# The perceived assortativity of social networks: Methodological problems and solutions

David N. Fisher[1], Matthew J. Silk[2] and Daniel W. Franks[3]

## Abstract

Networks describe a range of social, biological and technical phenomena. An important property of a network is its degree correlation or assortativity, describing how nodes in the network associate based on their number of connections. Social networks are typically thought to be distinct from other networks in being assortative (possessing positive degree correlations); well-connected individuals associate with other well-connected individuals, and poorly-connected individuals associate with each other. We review the evidence for this in the literature and find that, while social networks are more assortative than non-social networks, only when they are built using group-based methods do they tend to be positively assortative. Non-social networks tend to be disassortative. We go on to show that connecting individuals due to shared membership of a group, a commonly used method, biases towards assortativity unless a large enough number of censuses of the network are taken. We present a number of solutions to overcoming this bias by drawing on advances in sociological and biological fields. Adoption of these methods across all fields can greatly enhance our understanding of social networks and networks in general.

1. Department of Integrative Biology, University of Guelph, Guelph, ON, N1G 2W1, Canada.
   Email: davidnfisher@hotmail.com (primary corresponding author)

2. Environment and Sustainability Institute, University of Exeter, Penryn Campus, Penryn, TR10 9FE, UK
   Email: matthewsilk@outlook.com

3. Department of Biology & Department of Computer Science, University of York, York, YO10 5DD, UK
   Email: daniel.franks@york.ac.uk

# Introduction

Network theory is a useful tool that can help us explain a range of social, biological and technical phenomena (Pastor-Satorras et al 2001; Girvan and Newman 2002; Krause et al 2007). For instance, network approaches have been used to investigate diverse topics such as the global political and social system (Snyder and Kick 1979; Nemeth and Smith 2010) to the formation of coalitions among-individuals (Kapferer 1969; Zachary 1977; Thurman 1979; Voelkl and Kasper 2009). Networks can be described using a number of local (related to the individual) and global (related to the whole network) measures. One important global measure is degree correlation or assortativity (we use the latter term for brevity), which was formally defined by Newman (2002), although Pastor-Satorras et al. (2001) had calculated an analogous measure previously. The assortativity of a network measures how the probability of a connection between two nodes (individuals) in a network depends on the degrees of those two nodes (the degree being the number of connections each node possesses). The measure quantifies whether those with many connections associate with others with many connections (assortative networks or networks with assortativity), or if "hubs" form where well connected individuals are connected to many individuals with few other connections (disassortative networks or networks with dissassortativity). If the tendency for nodes to be connected is independent of each other's degrees a network has neutral assortativity. Assortativity is calculated as the Pearson's correlation coefficient between the degrees of all pairs of connected nodes, and ranges from -1 to 1 (Newman 2002). The Pastor-Satorras method involves plotting the degree of each node against the mean degree of its neighbours, and judging the network assortatative if the slope is positive and disassortatative if the slope is negative (Pastor-Satorras et al 2001). In this article we will focus on the Newman measure, as it has been more commonly used by the scientific community, gives a coefficient bounded between -1 and 1 rather than a slope, and typically is supported by a statistical test, something general absent from the reporting of the Pastor-Satorras method.

Assortativity is a key property to consider when understanding how networks function, especially when considering social networks. A network's robustness to attacks is increased if it is assortative (Jing et al 2007; Hasegawa et al 2012). However, the speed of information transfer and ability to act in synchrony is increased in disassortative networks (Di Bernado et al 2007; Gallos et al 2008). Thus the assortativity of a social network in which an individual is embedded can have a substantial impact on that individual.

Networks are typically found to be neutrally or negatively assortative (Newman 2002; Newman and Park 2003; Whitney and Alderson 2008). When considering all networks comprised of eight nodes, Estrada (2011) observed that only 8% of over 11,000 possible networks were assortative. Despite this general trend, social networks are often said to differ from other networks by being assortative (Newman 2003; Newman and Park 2003). This has led to those finding disassortativity in networks of online interactions (Holme et al 2004), mythical stories (Mac Carron and Kenna 2012) or networks of dolphin interactions (Lusseau and Newman 2004) to suggest they are different to typical human social networks. While the generality of assortativities in social networks has been questioned (Whitney and Alderson 2008; Hu and Wang 2009), a wide variety of recent research still states that this is a property typical of social networks (e.g. (Estrada 2011; Palathingal and Chirayath 2012; Mac Carron and Kenna 2013; Litvak and van der Hofstad 2013; Thedchanamoorthy et al 2014; Araújo et al 2014; Furtenbacher et al 2014)). With assortativity being a key network property, and subject of interest in a range of fields, this topic requires clarification.

In this paper we review assortativity in the networks literature, with emphasis on social networks. We assess the generality of the hypothesis that social networks tend to be distinct from other kinds of networks in their assortativities and explore whether the precise method of social network construction influences this metric. We go on to show how particular methodologies of social network construction could result in falsely assortative networks, and present a number of solutions to this by drawing on advances in sociological and biological fields.

1. Assortativity in social and other networks

Random networks should be neutrally assortative (Newman 2002). However, simulations by Franks et al. (2009) showed that random social networks constructed using group-based methods are assortative unless extensive sampling is carried out. Group-based methods are where links are formed between individuals not when they directly interact, but when they both are found in the same group, or contribute to a joint piece of work e.g. co-author a paper or appear in a band together. We hypothesised that the suggestion that social networks possess assortativity was due to a preponderance of social networks constructed using group-based methods in the early literature. We thus conducted a literature search, recording the (dis)assortativity of networks, network types (social or non-social) and for the social networks, method of construction (direct interactions of group-based). We expect that social networks built using group-based methods would be more assortative than the other two classes of network, which would be similarly assortative.

1.1 Literature search - Method

Assortativity has been calculated for a wide range of networks in the last decade. We compiled a list of the assortativity of published networks based on the table in Whitney &

Alderson (Whitney and Alderson 2008), literature searches with the terms "degree correlation" and "assortativity", and examining the articles citing Newman (2002). If it could not be determined how the network was constructed, or how large it was, the network was excluded. We did not include the average assortativity when reported from a range of similar networks when the individual scores were not reported. We also did not include assortativities of networks from studies re-analysing existing datasets, to avoid pseudo-replication. Only undirected networks were considered; see Piraveenan et al. (2012) for a review of assortativities in directed networks. We then classified these networks as non-social networks, social networks constructed using direct interactions, or social networks constructed using group-based methods. We then compared assortativity across network classes using a Kruskal-Wallis test as assortativity scores were not normally distributed. If this revealed significant differences among network classes, we then compared network classes to each other using Wilcoxon rank sum tests and to the a neutral assortativity of zero with Wilcoxon signed rank tests.

1.2 Literature search - Results

In published papers we found assortativities for 88 networks that met our criteria for inclusion, see Table 1. 52 of these were social networks, of which 25 were constructed using group-based methods. The assortativities for the network classes are shown as boxplots in Fig. 1. The Kruskal-Wallis tests indicated that there were differences among groups (Kruskal-Wallis $\chi^2 = 26.8$, d.f. = 2, $p < 0.001$). All networks types were different from each other, with the group-based social networks being more assortative than both other classes of network, and the direct social networks being more assortative than the non-social networks (all Wilcoxon rank sum tests, group based social networks vs. direct social networks, W = 214, n(direct) = 27, n(group-based) = 25, p = 0.024; group based social networks vs. non-social networks, W = 783, n(group-based social networks) = 25, n(non-social network) = 36, p <

0.001; direct social networks vs. non-social network, W = 716, n(direct social networks = 27), n(non-social network) = 36, p = 0.001). The direct social networks possessed neutral assortativities (mean = 0.054, Wilcoxon signed rank test, V = 213, n = 27, p =0.572), while the group-based social networks were assortative (mean = 0.157, Wilcoxon signed rank test, V = 288, n = 25, p < 0.001) and the non-social networks were disassortative (mean = -0.117, Wilcoxon signed rank test, V = 94.5, n = 36, p < 0.001).

**Table 1** 88 networks of various types and methods of construction. Type indicates the nature of the links between nodes, social are networks based on social interactions, technical are networks of interacting technology systems, biological are networks of some kind of biological process, transport networks are networks of a mode of transport in a particular area, mechanical networks are based on connections between parts in a defined object and links in commercial networks are formed when an employee moved from one company to another. Method of construction for social networks can be either "direct" where one to one interactions are used to build the network, or "group" where interactions are inferred based on shared use of physical space or contribution to a piece of work. All quoted degree correlations are from the Newman (2002) method of calculating assortativity.

| Network | Size | Type | Assortativity | Method of construction | Source |
|---|---|---|---|---|---|
| Beowulf (myth) | 74 | social | -0.1 | direct | (Mac Carron and Kenna 2012) |
| Cyworld (online) | 12048186 | social | -0.13 | direct | (Hu and Wang 2009) |
| Email address books | 16881 | social | 0.092 | direct | (Newman 2003) |
| Epinions neg (online) | 131828 | social | -0.022 | direct | (Ciotti et al 2015) |
| Epinions pos (online) | 131828 | social | 0.217 | direct | (Ciotti et al 2015) |

| Name | Size | Type | Value | Mode | Reference |
|---|---|---|---|---|---|
| Facebook (online) | 721000000 | social | 0.226 | direct | (Ugander et al 2011) |
| Flickr (online) | 1846198 | social | 0.202 | direct | (Hu and Wang 2009) |
| Gnutella P2P (online) | 191679 | social | -0.109 | direct | (Hu and Wang 2009) |
| Ground squirrels | 65 | social | 0.82 | direct | (Manno 2008) |
| Iliad (myth) | 716 | social | -0.08 | direct | (Mac Carron and Kenna 2012) |
| LiveJournal (online) | 5284457 | social | 0.179 | direct | (Hu and Wang 2009) |
| Mixi (online) | 360802 | social | 0.122 | direct | (Hu and Wang 2009) |
| MySpace (online) | 100000 | social | 0.02 | direct | (Hu and Wang 2009) |
| Nioki (online) | 20259 | social | -0.13 | direct | (Holme et al 2004) |
| Orkut (online) | 100000 | social | 0.31 | direct | (Hu and Wang 2009) |
| Pussokram (online) | 29341 | social | -0.048 | direct | (Holme et al 2004) |
| Slashdot neg (online) | 82144 | social | -0.114 | direct | (Ciotti et al 2015) |
| Slashdot pos (online) | 82144 | social | 0.162 | direct | (Ciotti et al 2015) |
| Student relationships | 573 | social | -0.029 | direct | (Newman 2003) |
| Táin (myth) | 404 | social | -0.33 | direct | (Mac Carron and Kenna 2012) |
| Twitter (online) | 4317000 | social | -0.025 | direct | (Wang et al 2014) |
| Whisper (online) | 690000 | social | -0.011 | direct | (Wang et al 2014) |
| Xiaonei (online) | 396836 | social | -0.0036 | direct | (Hu and Wang 2009) |
| YouTube (online) | 1157827 | social | -0.033 | direct | (Hu and Wang 2009) |
| Chinese science citations | 81 | social | -0.036 | direct | (Shan et al 2014) |
| Barbary macaque (grooming) | 141 | social | 0.351 | Direct | (Sosa 2014) |
| GitHub (online) | 671751 | social | -0.0386 | direct | (Lima et al 2014) |
| Australian dolphins | 117 | social | 0.003 | group | (Wiszniewski et al 2010) |
| Biology co-authors | 1520251 | social | 0.13 | group | (Newman 2002) |
| Birds (mixed species) | 93 | social | 0.29 | group | (Farine et al 2015) |
| Company directors | 7673 | social | 0.28 | group | (Newman 2002) |

| Name | Size | Type | Assortativity | Category | Reference |
|---|---|---|---|---|---|
| Condensed matter co-authors | 36458 | social | 0.18 | group | (Illenberger and Flötteröd 2012) |
| Condensed matter co-authors 1995-9 | 16729 | social | 0.18 | group | (Kossinets 2006) |
| Film actors | 449913 | social | 0.21 | group | (Newman 2003) |
| Killer whales | 7 | social | -0.48 | group | (Whitehead 2008) |
| New Zealand dolphins | 64 | social | -0.044 | group | (Lusseau and Newman 2004) |
| Maths co-authors | 253339 | social | 0.12 | group | (Newman 2002) |
| Physics co-authors | 52909 | social | 0.36 | group | (Newman 2002) |
| Physics co-author2 | 16264 | social | 0.18 | group | (Newman 2002) |
| Scottish dolphins | 124 | social | 0.17 | group | (Lusseau et al 2006) |
| Sticklebacks | 94 | social | 0.66 | group | (Croft et al 2006) |
| TV series actor collaboration | 79663 | social | 0.53 | group | (Hu and Wang 2009) |
| Grant proposals (accepted) | 24181 | social | -0.1018 | group | (Tsouchnika and Argyrakis 2014) |
| Grant proposals (rejected) | 46567 | social | -0.1145 | group | (Tsouchnika and Argyrakis 2014) |
| Brazilian co-authorship (Humanities) | 74490 | social | 0.3737 | group | (Mena-Chalco et al 2014) |
| Brazilian co-authorship (Linguistics, letters & arts) | 15375 | social | 0.3761 | group | (Mena-Chalco et al 2014) |
| Brazilian co-authorship (Engineering) | 15375 | social | 0.0273 | group | (Mena-Chalco et al 2014) |
| Brazilian co-authorship (Agricultural science) | 55695 | social | 0.0769 | group | (Mena-Chalco et al 2014) |
| Brazilian co-authorship (Biological science) | 75304 | social | 0.1404 | group | (Mena-Chalco et al 2014) |

| Name | Size | Type | Value | Category | Reference |
|---|---|---|---|---|---|
| Brazilian co-authorship (Exact and earth sciences) | 65221 | social | 0.1173 | group | (Mena-Chalco et al 2014) |
| Brazilian co-authorship (applied social sciences) | 48340 | social | 0.1373 | group | (Mena-Chalco et al 2014) |
| Brazilian co-authorship (health sciences) | 114169 | social | 0.1301 | group | (Mena-Chalco et al 2014) |
| Airports in Pakistan | 35 | transport | -0.47 | NA | (Mohman et al 2015) |
| Berlin U & S Bahn | 75 | transport | 0.096 | NA | (Whitney and Alderson 2008) |
| Bike | 131 | mechanical | -0.2 | NA | (Whitney and Alderson 2008) |
| Brain connections | 160 | biological | 0.058 | NA | (Im et al 2014) |
| Brain connections, Polymicrogyria | 160 | biological | 0.044 | NA | (Im et al 2014) |
| Car door | 649 | mechanical | -0.16 | NA | (Whitney and Alderson 2008) |
| Domain network (online) | 11174 | technological | -0.17 | NA | (Lee et al 2015) |
| Film references | 40008 | commercial | -0.057 | NA | (Spitz and Horvát 2014) |
| Fresh water food web | 92 | biological | -0.33 | NA | (Newman 2003) |
| Grand piano action key 1 | 71 | mechanical | -0.32 | NA | (Whitney and Alderson 2008) |
| Internet | 10697 | technical | -0.19 | NA | (Newman 2003) |
| Jet engine | 60 | mechanical | -0.13 | NA | (Whitney and Alderson 2008) |
| London underground | 92 | transport | 0.01 | NA | (Whitney and Alderson 2008) |
| Marine food web | 134 | biological | -0.26 | NA | (Newman 2003) |
| Metabolic pathways | 765 | biological | -0.24 | NA | (Newman 2003) |
| Moscow subway | 51 | transport | 0.18 | NA | (Whitney and Alderson 2008) |

| Name | Size | Type | Value | | Reference |
|---|---|---|---|---|---|
| Moscow subway & regional rail | 129 | transport | 0.26 | NA | (Whitney and Alderson 2008) |
| Neural pathways | 307 | biological | -0.23 | NA | (Newman 2003) |
| Power grid | 4941 | technical | -0.003 | NA | (Newman 2003) |
| Proteins | 2115 | biological | -0.16 | NA | (Newman 2003) |
| Pseudomonas strains | 37 | biological | -0.553 | NA | (Aguirre-von-Wobeser et al 2014) |
| Router network | 228298 | technological | -0.01 | NA | (Lee et al 2015) |
| Six speed transmission | 143 | mechanical | -0.18 | NA | (Whitney and Alderson 2008) |
| Software | 3162 | technical | -0.016 | NA | (Newman 2003) |
| Tokyo regional rail | 147 | transport | -0.09 | NA | (Whitney and Alderson 2008) |
| Tokyo regional rail & subway | 191 | transport | 0.043 | NA | (Whitney and Alderson 2008) |
| V8 engine | 243 | mechanical | -0.27 | NA | (Whitney and Alderson 2008) |
| World Wide Web | 269504 | technical | -0.067 | NA | (Newman 2003) |
| Yeast genes | 333 | biological | -0.15 | NA | (Lee et al 2015) |
| Yeast proteins | 1066 | biological | -0.12 | NA | (Lee et al 2015) |
| Copenhagen streets | 1637 | transport | -0.07 | NA | (Jiang et al 2014) |
| London streets | 3010 | transport | -0.06 | NA | (Jiang et al 2014) |
| Paris streets | 4501 | transport | -0.06 | NA | (Jiang et al 2014) |
| Manhattan streets | 1046 | transport | -0.26 | NA | (Jiang et al 2014) |
| San Francisco streets | 3110 | transport | -0.01 | NA | (Jiang et al 2014) |
| Toronto streets | 2599 | transport | -0.06 | NA | (Jiang et al 2014) |

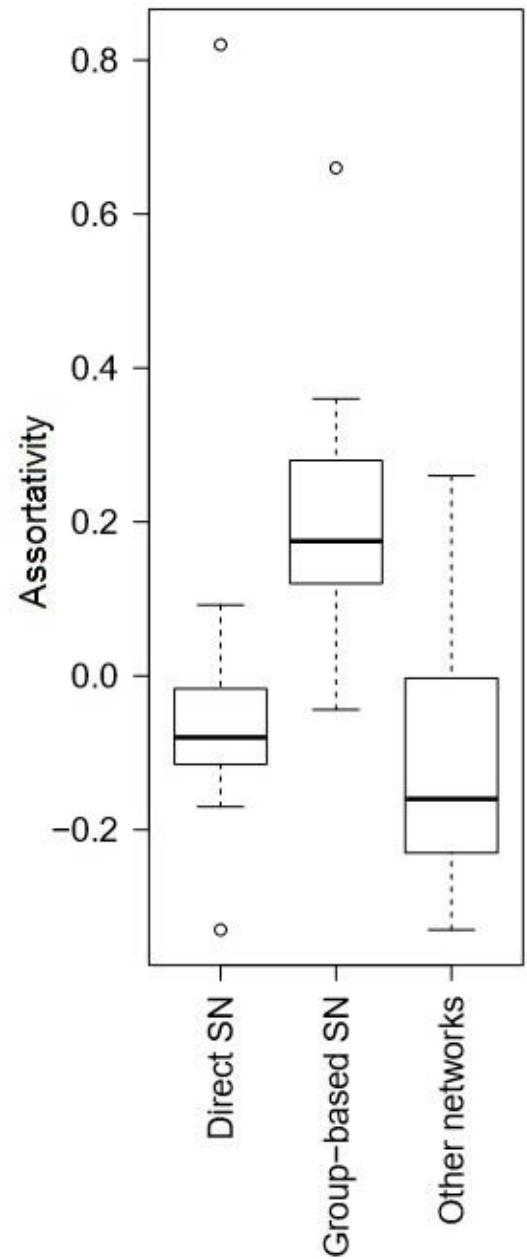

**Figure 1** Boxplots of assortativities in each class of network. Differences between all classes were statistically significant (all Wilcoxon rank sum tests, p < 0.025 in all cases). Direct social networks did not possess assortativities different from zero (mean = 0.054, p = 0.572), while the group-based social networks were assortative (mean = 0.157, p < 0.001) and the non-social networks were dissortative (mean = -0.117, p < 0.001; all Wilcoxon signed-rank tests).

### 1.3 Literature search - Conclusions

To confirm our hypothesis, that a preponderance of group-based methods makes it appear as if social networks typically possess assortativity, we found that group-based social networks were more assortative than direct social networks and non-social networks. However, direct social networks were still more assortative than non-social networks, which showed disassortativity on average. This therefore indicates that social networks only tend to be assortative if they are constructed with group-based methods, and non-social networks are typically disassortative. The processes that create dissortative networks with is a large and active research area (e.g. Yang 2014; Mussmann et al. 2014) and so we will not investigate this result further here. We also note that many of the direct social networks in our table were based on interactions online. This is perhaps problematic if interactions online are fundamentally different to human interactions in the real world. Our choice of networks was simply based on what was available. Therefore, to confirm whether the assortativities of offline and online human social networks made using the same method (i.e. with direct interactions) are different, more offline human networks using direct interactions need to be constructed and published.

The bias towards assortativity when using group-based methods has been used has previously been noted (Newman 2003; Croft et al 2008), but this appears to have escaped the notice of much of the research community, who continue to state that assortativity is a characteristic quality of a social network (e.g. (Estrada 2011; Palathingal and Chirayath 2012; Mac Carron and Kenna 2013; Litvak and van der Hofstad 2013; Thedchanamoorthy et al 2014; Araújo et al 2014; Furtenbacher et al 2014) but see (Whitney and Alderson 2008; Hu and Wang 2009)). This may be because no clear rationale for this has been presented, nor solutions offered to combat it. We now go on to present why we think this assortativity was being discovered with group-based methods, if they are erroneous, and how to avoid incorrect estimates of assortativity in the future.

## 2. Methodological pitfalls and false assortativity

In this section we will discuss the issues that exist when using assortativity to describe the structure of social networks. We explore how methodologies used to sample networks could influence their assortativities, and whether changing how assortativity is calculated may also be important. Understanding the consequences of the method used to sample networks on properties such as assortativity could have important implications for our understanding of networks in general, as well as contributing to the further development of relevant analytical techniques.

## 2.1 Group-based networks and assortativity

In situations where it is not possible or feasible to directly observe social interactions, social networks are constructed using group-based methods. This approach has been applied both in constructing collaboration networks in humans e.g. jazz musicians, scientific co-authorship networks and film actors (Watts and Strogatz 1998; Newman 2001; Gleiser and Danon 2003) and also in networks based on co-occurrence in a social group in animals e.g. song birds, dolphins and sharks (Lusseau 2003; Mourier et al 2012; Aplin et al 2012). Networks built using this method assume that every member of the group is associating with every other member of the group at each sampling census. This seems perfectly reasonable, hence this method for constructing social networks has been used by many studies (Newman 2004; Croft et al 2008). The assumption that meaningful social networks can be constructed based on this type of co-occurrence data has been termed "Gambit of the group" in the animal behaviour literature (Whitehead and Dufault 1999), which hints at the risk involved. This broad usage of group-based methods makes it very important that we fully understand the implications of using this method on the network and individual-level metrics calculated.

One study has previously investigated the influence of variation in sampling effort of group-based networks on various network-level metrics, including assortativity (Franks et al 2009). Franks et al. (2009) investigated the impact of different group-based sampling regimes, changing both the total number of censuses (the number of times a population is sampled using a group-based approach) and the proportions of individuals sampled at each census, in random networks. The assortativity of a random network, which should on average be zero (Newman 2002), was *always positive* if an insufficient number of censuses were completed before network construction (Franks et al. 2009; our emphasis). Crucially, any network not sampled intensively enough could show assortativity, either due to too few censuses or by not sampling enough of the population. Therefore, the use of group-based methods can produce a sampling bias, and requires further statistical analysis to determine the importance of the results obtained.

2.2 Modelling group-based sampling

We wanted to extend the findings of Franks et al. (2009) to demonstrate how assortativity changes as individuals are recorded in increasing numbers of groups. The aim was to show that in a system where freely-moving individuals form social groups randomly (as opposed to a system based on random social networks), assortativity would decline as the effect of two individuals being seen in the same group diminished. We used a simulation-based approach where individuals associated randomly to construct networks using a group-based approach in a simple population of 100. Individuals were allocated randomly between 20 possible groups during each "census" using a symmetric Dirichlet distribution (with a uniform shape parameter) to define the size of each group. This enabled us to generate variation around a fixed expected group size, whilst maintaining a fairly consistent number of groups in each census (occasionally the size of a group could be zero). Any remaining individuals were allocated to a random group, meaning that all individuals were sampled in each census. Association data was collected over 20 censuses, with this process repeated for 10 different repeats of the simulation. After each census cumulative association data was recorded and the networks were dichotomised to create binary networks. In these networks individuals are either connected if they were observed in the same group at least once, or not connected if never observed together. This is necessary as assortativity is an unweighted measure i.e. only the number of different associations an individual had are counted, not the frequency with which it associated with them. We discuss issues related to this method later. We then calculated assortativity for each set of cumulative association data. Simulations were carried out and degree correlations calculated in R 3.0.1 (http://www.R-project.org).

We plot the results of these simulations in Fig. 2. From this plot it is clear that networks possess assortativity at a low number of censuses, and become gradually more neutrally assortative as the number of censuses increases (Fig. 2). This pattern emerges because of how group-based approaches are used in network construction. At low numbers of censuses, individuals that have been found in the same groups will both be connected and have similar degree, thus giving the network assortativity (Newman 2003; Newman and Park 2003). As the number of censuses increases however, this connection will gradually break down. Therefore, our simple simulation model shows that the assortativity found using group-based sampling approaches is highly dependent on the number of censuses completed. Indeed a low number of censuses can lead to many network measures being distorted (Franks et al 2009; Perreault 2010). This is therefore an important consideration when deciding the sampling regimes used when constructing social networks using this method. The significance of the assortativity can only be established by comparison to appropriate null networks that truly highlight what aspects of the real network are interesting. This is achieved by randomly resampling observations using the correct group-size distribution (Bejder et al 1998), a method that will be outlined in more detail in section 3.2.

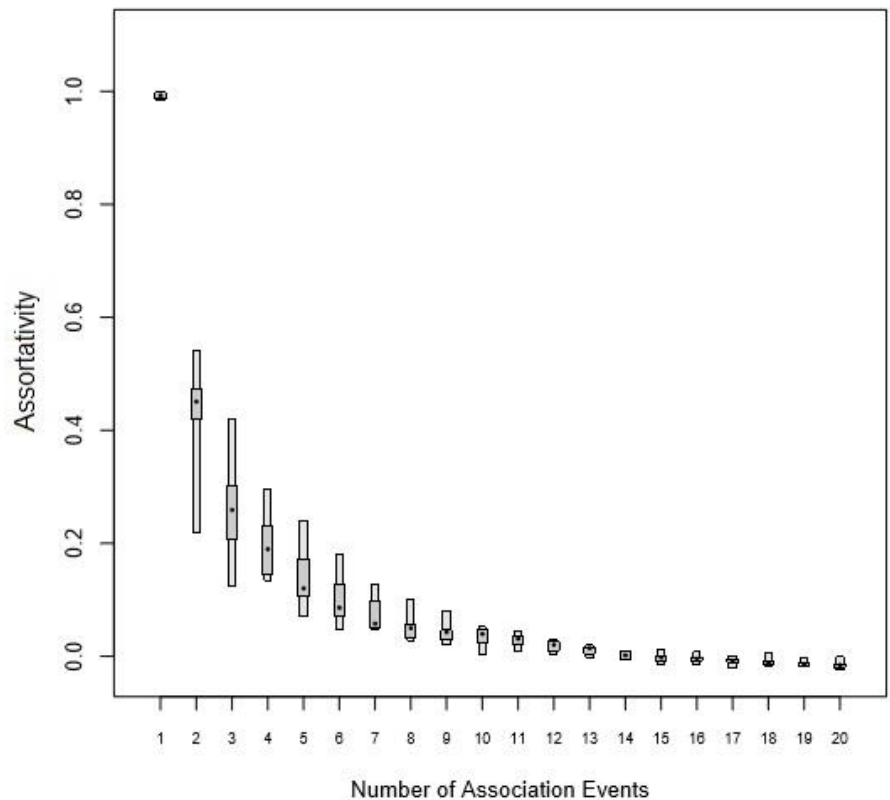

**Figure 2** The relationship between sampling effort (in terms of number of association events observed) and assortativity as a result of random social interactions in a simulated population. For each census the black point represents the median value, the dark grey box the interquartile range and the light grey box the range from 10 runs of the simulation.

2.2 Filtering networks

Filtering networks involves taking a weighted (or valued) network, which contains information about how strong associations are as well as whether they are present or absent, and removing edges below a certain weight i.e. infrequent or unimportant associations. This results in a binary network, with only edges above a certain weight present. The effect of dichotomizing networks to transform them from weighted to binary has been studied for a number of network-level metrics (Franks et al 2009). The effect of this on assortativity is striking. While there was a limited effect of filtering at a low threshold (and therefore removing few edges), filtering at a high threshold (removing relatively many edges) had a considerable impact on the assortativity even when a high number of censuses on the simulated networks was completed. While for unfiltered, weighted random networks the assortativity reduced to zero as expected when a sufficient number of censuses were completed, this was not the case for networks filtered with a high threshold. These networks typically remained assortative at all levels of sampling (Fig. 3 of Franks et al. 2009), despite being based on networks that were originally random and would be expected to neutrally assortative. For example, removing all edges with a weight of less than 0.5 (out of 1) meant that the assortativity reached 0.4 after 10 censuses, compared to 0.1 after 10 censuses if the threshold weight was set at 0.2. Many studies that have used group-based approaches to construct social networks have filtered the social networks produced before analysis (Croft et al 2008), and it seems likely that the use of this approach may have played an additional role in inflating the degree correlation calculated for social networks.

## 3. Solutions

There are a number of solutions available to reduce the occurrence of erroneous assortativity that address each of the major issues outlined above. These methods have been developed in different parts of the social network literature, and if used collectively can greatly improve our understanding of the true variation in assortativity in all kinds of social networks.

### 3.1 Increased sampling

When using group-based approaches, increasing the number of censuses above a threshold should produce a more accurate measure of the network's true assortativity (Franks et al 2009). Thus a suitable minimum number of censuses completed is required whenever using group-based approaches. If 80% of the population of interest can be sampled, then 10 censuses should be sufficient information on social interactions to accurately calculate network metrics. If only 40% of the population can be sampled, then 15-20 censuses may be required (Franks et al 2009). Collaboration networks, a key example of social networks that possess assortativity, typically do not have sampling periods. Instead all papers published over an agreed time period are looked at. To allow better comparison with networks and techniques that do have sampling periods, collaborations could be examined over particular, regular time periods, then each time period used as an "observation" from which to generate associations and so networks. However, if altering the sampling regime is not feasible, ensuring erroneous conclusions are not reached requires more rigorous statistical analysis.

### 3.2 Use of null models

Socio-biologists typically use resampling procedures to account for group-based sampling (and other, random) effects in their animal social networks (Bejder et al 1998; Whitehead and Dufault 1999). Resampling the observations used to construct the observed network using the original group-size data, and potentially additional biologically meaningful constraints, provides a large number of null networks against which to compare the observed network and test relevant hypotheses (Bejder et al 1998). This method has been continually revised to take into account the non-independence of group sightings (Sundaresan et al 2009) and the risk of producing biased group compositions (Krause et al 2009). This approach has been used in a number of studies (e.g. Wey et al. 2013; Aplin et al. 2013) and means that assortativity would only be considered interesting if the null model does not also display assortativity. We found one example of this approach being used in human social networks, in a study performed on a network of a board of directors (Newman and Park 2003). However, no statistical tests were used for this comparison, and the method does not seem to have been universally adopted by the wider social network community. The importance of adopting this method is further highlighted the results of Franks et al. (2009) which show that assortativity can be expected in networks where individuals randomly interact.

Random permutations are used by other network analysis techniques such as quadratic assignment procedure (QAP) to overcome problems with structural autocorrelation (Krackhardt 1988), but the Bejder et al. (1998) method and extensions go further by randomly permuting the raw data, rather than the network matrix, to account for the distribution of group sizes and any other biases in sampling. Appropriate null models like these are the most effective way to make statistical inferences about networks constructed using group-based approaches, where controlling for effects such as group-size is of considerable importance. There have been some recent examples of appropriate null models being used to test various hypotheses in the wider social networks literature (e.g. (Hanhijarvi et al 2009; La Fond and Neville 2010)) and we add to their calls for this to be more widely embraced.

### 3.3 Analysing weighted networks

Weak interactions are potentially important (e.g. Granovetter 1973; Farine 2014), and removing them can further increase error in the degree correlations calculated (Kossinets 2006; Franks et al 2009). If a population is observed for an extended period of time, and a large number of censuses are performed, the probability that two individuals are never observed to associate approaches zero. In a binary network, this would result in all individuals having the same, maximum, degree (which would be the population size -1), and therefore the assortativity would increase before becoming undefined. Therefore, it is preferable to use weighted, unfiltered networks. For example, if collaboration networks are based on some measure of the information contained in an email (Garton et al 2006), or the relative roles of those involved (Rowe et al 2007), you could then produce weighted networks that would be more informative to study and be less likely to be falsely assortative. The analysis of weighted networks is becoming increasingly manageable (e.g. Opsahl et al 2010), reducing the need to filter networks and analyse them as binary data. As such, this approach should only be used when absolutely required (Noldus and Van Mieghem 2015), preferably with filtering using low thresholds (few edges removed). The filtering of some networks in unavoidable, for example in brain imaging data which is likely to be very noisy (Iturria-Medina et al 2007), but we still recommend the lowest amount of filtering possible while accounting for edge uncertainty. Additionally, it would also be highly beneficial to continue to develop social network analysis in weighted networks to further reduce the requirement for binary network data, for example by continuing the development of exponential random graph models that can be used in weighted networks (Krivitsky 2012; Krivitsky 2015).

### 3.4 Using diadic over group-based approaches

The use of group-based methods is instrumental in creating some of the problems we suggest solutions for in previous subsections. If possible, alternative methods to group-based approaches should be used when constructing a network. For example, instead of paper co-authorship being used to define interactions between scientists, direct, diadic interactions based on correspondence such as emails could be used (Garton et al 2006; Rowe et al 2007). Human social networks based on communication data such as with mobile phones or online are also examples of diadic interactions, and are becoming increasingly common (Garton et al 2006; Hu and Wang 2009; Choudhury and Mason 2010; Peruani and Tabourier 2011; Expert et al 2011). In animal networks, this can be achieved by focussing on observation of suitable behavioural interactions (Manno 2008; Wey and Blumstein 2010) or using reality mining methodologies (Krause et al 2013), which are more fully outlined in section 3.5. Social networks constructed using direct interactions on average possess neutral assortativity (see above), indicating they are not prone to the same problems with false assortativities that networks built using group-based approaches are.

### 3.5 Modern technology

Missing associations, and therefore edges in the network, is a problem largely unique to social networks. This can be prevented in human social networks, especially collaboration networks, by examining information-rich communications data. Information rich interpersonal communications such as texts, tweets and emails can be used to construct networks on a very large scale with a high degree of accuracy (Choudhury and Mason 2010). In animal social networks a solution is to use reality mining (Krause et al 2013), the concept of which has been borrowed from the sociology social network literature (Eagle and Pentland 2005). In these studies, modern technologies such as GPS trackers or proximity loggers are used to track animal movements and monitor interactions or accurately infer associations. For example, proximity loggers have been used to automatically record associations based on spatial and temporal overlap between individuals in animals such as cows, Tasmanian devils, crows and badgers (Böhm et al 2009; Hamede et al 2009; Rutz et al 2012), although see (Drewe et al 2012; Boyland et al 2013) for discussion of the potential problems associated with using these devices.

These methods substantially reduce the number of missing edges in the network. Additionally, even when they don't allow us to move towards constructing networks using direct interactions, they can make it much more feasible to complete a large number of censuses if a group-based approach is still required. Furthermore, with careful thought technologies such as these could be used to complement the study of human social networks, potentially combined with data gathered using other methods. For instance, in the work place, are those an employee physically associates with e.g. at the water cooler or on a cigarette break, the same as those they communicate with electronically? What about different types of electronic communication, e.g. Facebook messages compared to work emails? Such application of modern technologies will only enhance the ability of scientists to measure social interactions in a wide range of networks.

### 3.6 Alternatives to the Newman degree correlation measure

Recent research has suggested that calculating the correlation coefficient proposed by Newman (2002) may not be appropriate for large networks, as it tends to mis-estimate the assortativity, especially in disassortative networks (Litvak and van der Hofstad 2013). Litvak and van der Hofstad (2013) suggested that using a method that ranks the degree of nodes, rather than using their absolute values (like a Spearman's rank correlation coefficient), may produce more valid results. Using this measure, the assortativity was consistent across a wide range of network sizes and typically consistently different from those calculated using Newman's assortativity coefficient (Fig. 1 of Litvak and van der Hofstad 2013). It may be that changing the way that assortativity is calculated could reduce the differences between group-based networks, other social networks, and different network types. This can be combined with the Bejder et al. (1998) method to control for both group size variation and biases associated with whole network size.

Alternatively, several authors have proposed individual-orientated metrics to quantify the tendency for well-connected individuals to connect to other individuals. The "Rich-clubs" of Zhou and Mondragon (Zhou and Mondragón 2004; Colizza et al 2006), and the number of "differences" in node degree between neighbours of Thedchanamoorthy et al. (2014) may both be robust to these pitfalls. Understanding the consequences of using different statistical approaches for calculating degree correlations should be the subject of further modelling work, in order to determine whether this can have an important influence to the properties with which a network is prescribed.

### Conclusions

Assortativity has often been suggested to be a property of social networks that makes them different from non-social networks. We show that it is likely that this phenomenon may be driven by the data collection methods, resulting in previous studies overestimating the extent to which social networks possess assortativity. With alterations to methods or analyses however, this problem with erroneous assortativities being described can be mitigated.

Through the use of methods initially developed in biology and sociology, we suggest that careful application of sampling and statistics can afford us increased confidence that finding assortativity in social networks is genuine, and not an artefact of the methodology used to construct them. It may well be that social networks are unique amongst types of network; there are certainly good reasons why they could have developed to be more assortative than non-social networks e.g. increased robustness (Jing et al 2007; Hasegawa et al 2012). However, given that information transfer and the ability to act in synchrony are greater in networks that show disassortativity (Di Bernado et al 2007; Gallos et al 2008), it would perhaps be surprising if this property is as universal as has previously been described (Newman 2002; Newman and Park 2003).

Many of the methods discussed have been developed in one part of the social networks literature, and would benefit researchers in other fields greatly. Furthermore, following the integration of methods used in these different fields, it is likely that new insights will become available and further progress made in our ability to make inferences about assortativity in networks more generally.